\documentstyle[aps,epsfig]{revtex}
\begin{document}
\draft
\title{ Spin in a Fluctuating Field: The Bose($+$Fermi) Kondo models}
\author{Anirvan M. Sengupta} 
\address{Serin Physics Laboratory, Rutgers University, Piscataway, NJ 08854}
%\address{and}
%\address{Raman Research Institute, Bangalore 560080}
\maketitle
\begin{abstract}
I consider  models with an impurity spin coupled to a fluctuating 
gaussian field with or without additional Kondo coupling of the conventional 
sort. In the case of isotropic fluctuations, the
renormalisation group flows for these models have controlled
fixed points when the autocorrelation of the gaussian field $h(t)$,
$<T h(t) h(0)> \sim {1\over t^{2-\epsilon}}$ with small positive $\epsilon$.
In absence of any additional Kondo coupling, I  get powerlaw decay of spin
correlators, $<T S(t) S(0)> \sim {1\over t^{\epsilon}}$ . For negative
$\epsilon$, the  spin autocorrelation is constant in long time limit.
The results agree with calculations in Schwinger Boson mean field theory.  
In presence of a Kondo coupling to itinerant electrons, the model shows
a phase transition from a Kondo  phase to a field fluctuation
dominated phase. These models are good starting points for understanding 
behaviour of impurities in a system near a zero-temperature magnetic transition.
They are also useful for understanding the dynamical local mean field
theory of Kondo lattice with  Heisenberg (spin-glass type) magnetic 
interactions.

\end{abstract}
\pacs{PACS numbers: 75.20.Hr, 75.30.Hx, 75.30.Mb}

Theory of  metallic spin glasses provides an example 
where Kondo effect and magnetic fluctuations compete on par at the
local mean field theory level\cite{larged}.
The local mean field theory of this model has an impurity spin coupled to 
a fluctuating gaussian Weiss field in addition to a Kondo coupling of the 
conventional sort. The treatment of metallic spin glass, so far,
has dealt with Kondo effect in a cavalier fashion \cite{spinglass,sg2}.
We do not have too much insight into the critical behaviour of the
impurity model discussed above. A good starting point would
be understanding the phase diagram of the impurity model without 	
the self-consistency condition  imposed by mean field theory 
of the spin-glass model. The problem of a spin coupled to a 
gaussian Weiss field which is power-law correlated in time,
with or without additional Kondo coupling gives rise to
a class of impurity models which are of interest on their own.

Let us first look at the  model without Kondo coupling.
Consider a spin $\vec{S}$ in a fluctuating magnetic field $\vec{h}(t)$. 
The hamiltonian is the following:

\begin{equation}
H=\gamma_z h^z S^z +\gamma_{\perp}(h^+ S^- +h^- S^+)+H_{bulk}(h) 
\end {equation}

$H_{bulk}$ is such that the fluctuation of $h$'s is gaussian. It is enough
to know the two point functions. For convenience we will work with Euclidean
time
\begin{equation}
<T_{\tau} h^a(\tau) h^b(0)>=D^{ab}(\tau)\sim \delta^{ab} /\tau^{(2-\epsilon)}
\end{equation}

I will call this model the Bose-Kondo model. Its behaviour is very
different from the usual Kondo model where the spin couples to spin
density of fermions. The spin density of fermions has non-trivial
three-point (and higher-point) functions which are crucially important
in the long time limit. I will be interested in the dynamics of the spin,
especially in the spin autocorrelation function.

This is a nontrivial problem. In a path integral formulation using
coherent state representation for the spin, one gets the effective
action
\begin{equation}
A_{eff}=\int\int d\tau_1 d\tau_2\sum_{ab}\gamma^a\gamma^b
 D^{ab}(\tau_1-\tau_2) S^a(\tau_1)S^b(\tau_2)
       +i\mbox{Berry Phase}[\vec{S}(\tau)]
\end{equation}
after integrating out the Weiss field. If one could ignore the Berry phase 
term, this would be a simple classical long-range spin model. We know
a lot about such model. For example, in the paramagnetic
phase, spin correlator have the same power law, as that of the long
range coupling. However, when the system is isotropic, i. e.
$\gamma^a\gamma^bD^{ab}(\tau)=\gamma^2D(\tau)\delta^{ab}$,
the Berry phase term changes the behaviour 
of the system completely. The spin autocorrelation
has a power law which is, generically, {\it different} from the
field-field correlator exponent.

Let us begin by considering the special case where $\epsilon=0$. 
Renormalisation of the coupling involves integrals of the form
$\int_0^{\infty} d\tau_2 \int_{-\infty}^0 d\tau_1 D^{ab}(\tau_2-\tau_1)$ which
are logarithmically divergent. The perturbative beta-functions for 
the couplings turn out to be

\begin{eqnarray}
\beta(\gamma_z)&=&-{1\over 2}\gamma_z \gamma_{\perp}^2 \\
\beta(\gamma_{\perp})&=&-{1\over 4}\gamma_{\perp}(\gamma_z^2 +\gamma_{\perp}^2)
\end{eqnarray}

\begin{center}
\begin{figure}[!b]
\epsfig{file=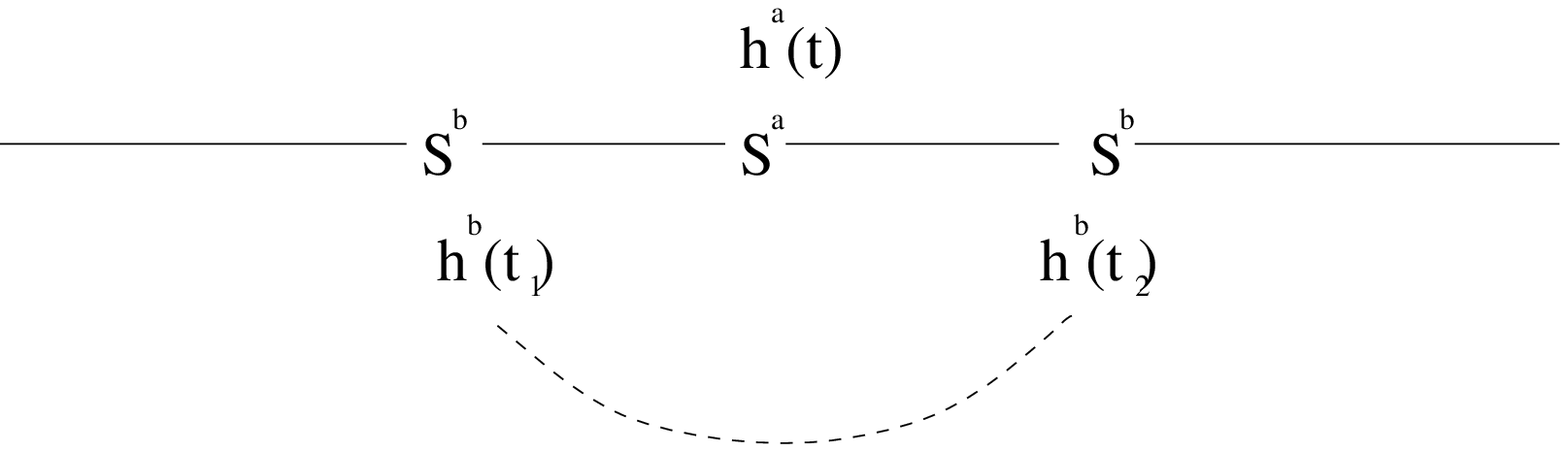,height=4cm,angle=270}\\
\caption{Renormalisation of $\gamma$.}
\end{figure}
\end{center}

From this point onward I discuss mostly the symmetric case, 
$\gamma_z=\gamma_{\perp}=\gamma$.

\begin{equation}
\beta(\gamma)=-{1\over 2}\gamma^3+o(\gamma^5)
\end{equation}

Now I consider small nonzero $\epsilon $. Dimension analysis of the coupling
$\gamma$ tells us that

\begin{equation}
\beta(\gamma)={\epsilon \over 2}  \gamma -{1\over 2} \gamma^3+\cdots
\end{equation}

This equation has a stable nontrivial fixed point at $\gamma_{\ast}^2=\epsilon$ 
when $\epsilon > 0$. Note that $\gamma \rightarrow -\gamma$ is a symmetry.
Hence, sometimes it is more useful to write the beta function in terms of
$g=\gamma^2$.
\begin{equation}
\beta(g)=\epsilon g - g^2 +\cdots
\end{equation}
To calculate the spin-spin auto correlation, we evaluate the correction
perturbatively in $\gamma_{\ast}$. Including  the  second order correction in
$\gamma$ to the spin autocorrelation, $ <T_{\tau} S(\tau) S(0)> \sim
1-\gamma_{\ast}^2 \log \tau$ leading to a power-law 
\begin{equation}
<T_{\tau} S(\tau) S(0)> \sim 1/\tau^{\epsilon}
\end{equation}
upto order $\epsilon$.

It is quite remarkable that  the Schwinger boson  calculation  gives
the same power-law, for arbitrary $\epsilon$. This calculation is
justified for spin belonging to certain representations of $SU(N)$
when $N \rightarrow \infty$. 
 The case $\epsilon=1$ appeared in the self-consistent
mean field theory of a gapless spin-fluid phase \cite{subir}.

The Bose-Kondo model has some interesting applications as an impurity
model. One of them is the problem of a non-magnetic site in an
antiferromagnet near a quantum critical point. The usual procedure
for deriving the long distance field theory for quantum antiferromagnets
is to take pairs of neighbouring sites and associate a staggered
magnetisation  order parameter with the pair. Then one writes down
the effective hamiltonian in terms of this order parameter. When there
is a non-magnetic site, there will be a pair with one magnetic
and one non-magnetic site. The corresponding effective hamiltonian would have a 
spin coupled at one site to the order parameter of a sigma-model,
the long distance theory of the anti-ferromagnet. This model may be
of use in studying effects of Zn doping in cuprates.

When the sigma model is in the the quantum critical region, analysing
this impurity model is hard. For dimensions near two, the order parameter
fluctuations are far from gaussian. However, for spatial dimension
$d=3-\epsilon$, $\epsilon$ small one can replace the sigma-model by
a $\phi^4$ field theory and use $\epsilon$ expansion to treat the interactions.
Upto order $\epsilon^2$, one can use the result of the gaussian approximation.
I believe this leads to impurity contributions to the susceptibility
which diverge like $1/T^{1-\epsilon}$ as a function of temperature
$T$ for small $\epsilon$. Since these spins are not very effectively
quenched, a small concentration of them can easily lead to a magnetic
state.

So much for the Bose-Kondo model with just a Gaussian Weiss field. 
What happens when it is in competition with the usual Kondo effect?
Let there be some additional fermionic degrees of freedom
interacting with the spin through antiferromagnetic Kondo coupling $J$.
Clearly, for $\epsilon >0$,$J$ is irrelevant, for small $J$.
For big enough Kondo coupling, there is a Kondo dominated phase
where $<T_{\tau} S(\tau) S(0)> \sim 1/\tau^2$.

The physics is captured by the beta functions:
\begin{eqnarray}
\beta(J)&=&-{1\over 2}\gamma^2 J +J^2 +\cdots\\
\beta(\gamma)&=&{\epsilon \over 2}\gamma -{1 \over 2} \gamma^3+\cdots
\end{eqnarray} 
The modefication of the magnetic coupling due to Kondo effect shows up
in the term of the order $-\gamma J^2$ in $\beta(\gamma)$.
 However, to determine the fixed points
to the lowest odrer in $\epsilon$, we don't need this.

In terms of $g=\gamma^2$
\begin{eqnarray}
\beta(J)&=&-{1\over 2}g J +J^2 +\cdots\\
\beta(g)&=&\epsilon g -g^2+\cdots
\end{eqnarray} 

\begin{center}
\begin{figure}[hbt]
\epsfig{file=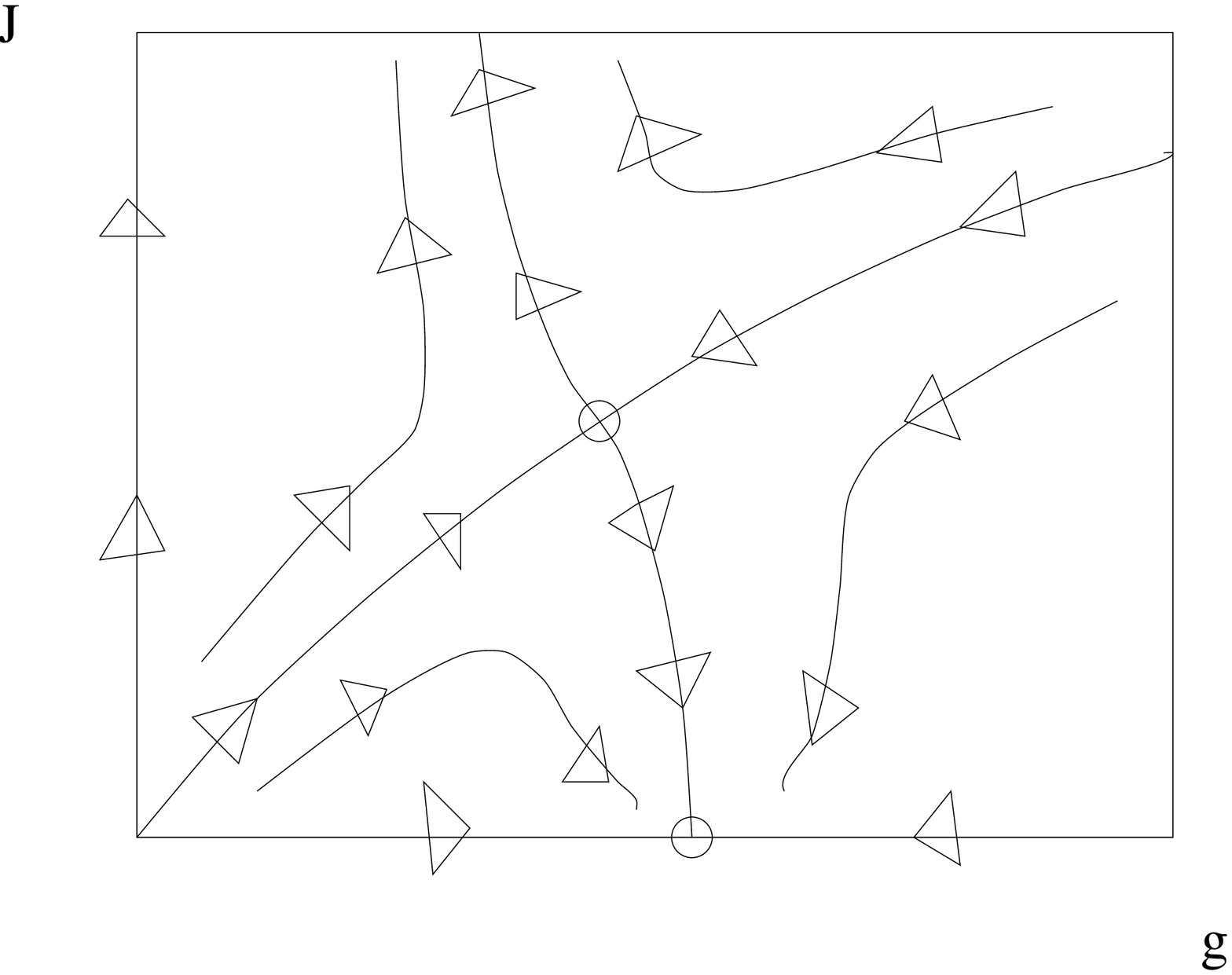,height=5cm,angle=270}\\
\caption{Renormalisation Group flow in the $g-J$ plane.}
\end{figure}
\end{center}

These equations have an unstable fixed point at
\begin{eqnarray}
g_{\ast}&=&\epsilon\\
J_{\ast}&=&{\epsilon \over 2}
\end{eqnarray}

Notice the line in the  $g-J$ plane which flows {\it into} the unstable
fixed point. Points above that line
flow to the strong Kondo fixed point. Points below the line flow to
the magnetic fixed point $J=0,g=\epsilon$.

At the unstable fixed point, spin autocorrelation ought
to have a different power law. However, up to order $\epsilon$
the exponent remains the same.

This  phase diagram can be obtained in a different limit, where
$\epsilon$ is arbitrary but the spin belongs to an antisymmetric
representation of $SU(N)$ with $N \rightarrow \infty$ \cite{gabi}. One can
treat the Kondo effect by slave boson mean field theory. In the
magnetic fluctuation dominated phase, the slave boson does not condense.

All the magnetic fixed points discussed so far, the magnetic fluctuation
dominated one as well as the fixed point at the border of the Kondo phase,
depend crucially on isotropy.
Anisotropy is a relevant perturbation around these fixed points. This is
another aspect of Bose-Kondo models which differs from the usual fermionic
Kondo model. In the usual Kondo model,  anisotropies in antiferromagnetic
couplings renormalise away. For the Bose-Kondo model with anisotropic
couplings, the biggest coupling, say $\gamma^z$, wins. One might as well 
ignore the other couplings, $\gamma^x$ and $\gamma^y$, in that case. 
Such problems have been discussed before \cite{Si} in a slightly different
context. 

In conclusion, I have studied the problem of a spin coupled to a fluctuating 
power-law correlated gaussian field, with or without additional Kondo coupling
of the conventional sort. These models have interesting phase transitions
and critical behaviour, when the magnetic fluctuations are isotropic. 
Apart from being interesting in their own right, such
impurity models provide a good starting point for understanding the
local mean field theory of the Kondo lattice with Heisenberg
spin-glass type magnetic interactions \cite{gabi}.

I acknowledge useful discussions with G. Kotliar, A. Millis, S. Sachdev,
Q. Si and C. M. Varma.  This work was supported by the grant NSF 
DMR 96-32294.

\end{document}